\documentclass [prd,twocolumn,nofootinbib,showpacs]  {revtex4}
\def\bbbone{{\mathchoice {\rm 1\mskip-4mu l} {\rm 1\mskip-4mu l}
{\rm 1\mskip-4.5mu l} {\rm 1\mskip-5mu l}}}
\begin{document}

\title{%
Non-commutative space-time of Doubly Special Relativity theories }
\author{ Jerzy \surname{Kowalski--Glikman}}
\email{jurekk@ift.uni.wroc.pl}
\thanks{Research  partially supported
by the    KBN grant 5PO3B05620.}
\author{Sebastian Nowak}
\email{pantera@ift.uni.wroc.pl}
\affiliation{ Institute for Theoretical
Physics\\ University of Wroc\l{}aw\\ Pl.\ Maxa Borna 9\\
Pl--50-204 Wroc\l{}aw, Poland}

\begin{abstract}
Doubly Special Relativity (DSR) theory is a recently proposed
theory with two observer-independent scales (of velocity and
mass), which is to  describe a kinematic structure underlining the
theory of Quantum  Gravity.  We  observe that there is infinitely
many  DSR constructions of the energy-momentum sector, each of
whose can be promoted to the $\kappa$-Poincar\'e quantum (Hopf)
algebra. Then we use the co-product of this algebra and the known
construction of $\kappa$-deformed phase space via Heisenberg
double  in order to derive the non-commutative space-time
structure and description of the whole of the DSR phase space.
Next we show that contrary to the ambiguous structure of the
energy momentum sector, the space-time of the DSR theory is unique
and equivalent to the theory with non-commutative space-time
proposed long ago by Snyder. This theory provides non-commutative
version of Minkowski space-time enjoying ordinary Lorentz
symmetry.  It turns out that when one builds a natural phase space
on this space-time, its intrinsic length parameter $\ell$ becomes
observer-independent.

\end{abstract}
\maketitle
\section{Introduction}

Most approaches to quantum gravity work in top-to-bottom fashion:
one tries to define a more or  less complete theory (like strings
theory or loop quantum gravity) and then to work out some picture
of low energy phenomena that can be in principle confronted with
experiment. This procedure is in a sharp contrast with the way the
General Theory of Relativity has been first formulated. Einstein
started with  Special Relativity, and then tried to remove the
tension between this theory and the Newtonian theory of gravity.
Such a bottom-to-top way is certainly easier because one starts
with relatively simple problems, and only after getting some
experience the more difficult ones are started being addressed.
However, contrary to the situation Einstein was facing at the
beginning of XX century, on our way to quantum gravity we are in
much more difficult situation having in our disposal virtually no
experimental input (see however \cite{gacphen} for review of the
current status of ``Quantum Gravity Phenomenology''.) The
development of the top-to-bottom theories provided us however with
some rather robust intuitions as to what the starting point of
bottom-to-top procedure should be. First of all quantum gravity
should certainly describe the flat geometry, i.e., the structure
of space-time when gravitational field is switched off. Second,
the top-to-bottom approaches seem to indicate the existence of a
fundamental scale, which could be identified with Planck scale (of
length or  mass.) These two observations form the basis for the
studies presented in this paper.

About a year ago, in his seminal papers \cite{gac1}, \cite{gac2},
\cite{gacnew} G. Amelino-Camelia   made the major breakthrough in
the bottom-to-top approach to quantum gravity. He observed  that
if the Planck scale  is treated not as a coupling constant, but as
an observer-independent scale, then we immediately find  ourselves
with direct conflict with basic principles of Special Relativity,
which teaches us that masses (and lengths) are not
observer-independent. To overcome this difficulty Amelino-Camelia
proposed to consider theory with two observer-independent scales,
dubbed by him the Doubly Special Relativity (DSR). He also
suggested that such a theory may have as an algebra of symmetries
some kind of a quantum (Hopf) algebra, which would be an extension
of the standard Poincar\'e  algebra. The existence of the second
observer-independent scale is in this approach a remnant of
Quantum Gravity, and thus Doubly Special Relativity is to be
viewed therefore as a Quantum Gravity in flat space-time, or, in
other words, Quantum Special Relativity.

It turned out \cite{jkgminl}, \cite{rbgacjkg} that the DSR
proposal can be realized in the  framework of the quantum
$\kappa$-Poincar\'e  algebra \cite{lunoruto}, \cite{maru}. This
algebra  contains a natural deformation parameter of dimension of
mass $\kappa$ by construction (along with the speed of light $c$
which we will put here equal $1$.) It has been explicitly showed
in \cite{rbgacjkg} that one of the basic physical prediction of
the DSR theory with $\kappa$-Poincar\'e  algebra  playing the role
of an algebra of symmetries is the existence of the fundamental
mass scale, which is observer-independent.

The quantum $\kappa$-Poincar\'e  algebra is in the same relation
to DSR theory as  the standard Poincar\'e  algebra is related to
Special Relativity. In other words DSR is a physical theory in
which the description of  energy-momentum sector is provided by
$\kappa$-Poincar\'e quantum algebra. However the knowledge of this
algebra is only the first step in construction of the DSR theory.
The reason is at least twofold.

First of all the $\kappa$-Poincar\'e  algebra is a quantum
nonlinear algebra, and thus,  contrary to the standard Poincar\'e
case there are no restrictions on nonlinear transformations among
momenta. This means that from the mathematical point of view
different bases of the algebra related by arbitrary (analytical)
transformations of momenta are completely equivalent. Up to date
four such bases have been investigated in details: the
``standard'' basis with deformed Lorentz sector \cite{lunoruto},
the bicrossproduct basis \cite{maru}, the classical basis
\cite{kolumaso}, and recently a new basis proposed by Magueijo and
Smolin \cite{JoaoLee}, \cite{juse}. It is therefore an open
question whether these mathematically equivalent bases are
equivalent physically, or if one of them is singled out by some
physical considerations.

Secondly, the $\kappa$-Poincar\'e quantum algebra is an algebra
describing in a direct way only the energy-momentum sector of the
DSR theory. The knowledge of this sector alone is insufficient to
consider many physical problems. Moreover, as argued in
\cite{lunoDSR} if one sticks only to the energy-momentum sector,
the resulting theory is based on a brute force, unjustified
deformation of the standard Poincar\'e algebra and thus not very
appealing. It is the co-algebra of the quantum algebra that
provides a way out of this dilemma. Indeed, as we show below  the
co-algebra can be used to derive the form of the commutator
algebra on the whole of the phase space. The $\kappa$-deformed
phase space  has been first obtained in \cite{crossalg},
\cite{luno}, \cite{lukps} for  bicrossproduct basis. This result
was extended  in \cite{juse} for other bases. However the emerging
picture of the phase space was still unclear. Here we want to
investigate the properties of this space in more details. In
particular, we will address the question which of these properties
are universal, independent of the energy-momentum basis one starts
with, and which are not. If indeed all the bases are physically
equivalent only the former may correspond to physical
``observables''.

The plan of this paper is as follows. In sections II and III we
show that under rather mild assumption an arbitrary
energy-momentum algebra forms, along with the undeformed Lorentz
algebra a Hopf algebra. This is a very promising result, since the
Hopf algebra structure makes it possible to extend the
energy-momentum algebra to the algebra of space-time. This
extension turns out to have a remarkable property, namely that the
form of space-time non-commutativity is independent of the
energy-momentum basis one starts with. This result is proved in
section III. Next, in section IV  we show that also the way how
positions transform under action of Lorentz generators is
energy-momentum basis independent. Thus the space-time sector of
DSR is perfectly unambiguous, and can therefore serve as a basis
to construct a physical theory. Moreover, it turns out that the
resulting theory is exactly the one proposed in the  paper of
Snyder \cite{snyder}, in which the idea of space-time
non-commutativity was spelled out for the first time. The natural
phase-space algebra associated with Snyder's non-commutative
Minkowski space-time is presented in section VI. Section VII is
defoted to conclusions and outlooks.

\section{Three bases of $\kappa$-Poincar\'e quantum algebra}

In this section we present general results describing some
algebras  furnishing the energy-momentum sector of DSR theory. Our
aim here is twofold: first we  present concrete examples, of which
we will make use in the following sections, and second we prove
statements concerning the general nature of $\kappa$-Poincar\'e
quantum algebras, and therefore applicable to any DSR theory.

The $\kappa$-Poincar\'e quantum algebra is a quantum (Hopf)
algebra being  a deformation of the Poincar\'e algebra of special
relativity. This algebra has been proposed first in the paper
\cite{lunoruto} (see also \cite{rurev} for early review). However
the algebra presented there has been written in the so-called
standard basis, which does not satisfy the natural requirement
that the action of the Lorentz sector integrates to a group. Only
few years later in the paper \cite{maru} the bicrossproduct basis
was introduced in which the Lorentz sector was undeformed.
Therefore, to distinguish these two cases we  introduce the notion
of $\kappa$-DSR algebra  which is an $\kappa$-deformed Heisenberg
algebra  on the phase space of the system with the basis
satisfying
\begin{enumerate}
\item the Lorentz sector of this algebra is undeformed,
\item the action of rotations  on momenta is classical\footnote{Let
us note that the deformation can be associated only with one
dimension and it follows from the requirement of rotational
symmetry that we choose this direction to be timelike. One should
remember however that there exists a different $\kappa$-Poincar\'e
theory in which one deforms the algebra along null direction.},
\item the space-time commutators and the ones between positions and
momenta are uniquely defined by the co-product and appropriate
pairing (see below), and \item in the limit when the deformation
parameter $\kappa \rightarrow\infty$ the algebra becomes the
classical phase space algebra, i.e, the Poincar\'e algebra along
with the standard canonical commutational relations between
positions and momenta (with the trivial co-algebra sector).
\end{enumerate}

To date three examples (or, better to say, bases) of $\kappa$-DSR
algebra have been analyzed, and for all of them, according to
postulates (a) and (b):
$$
[M_i, M_j] = i\, \epsilon_{ijk} M_k, \quad [M_i, N_j] = i\, \epsilon_{ijk} N_k,
$$
\begin{equation}\label{1}
  [N_i, N_j] = -i\, \epsilon_{ijk} M_k.
\end{equation}
and
\begin{equation}\label{2}
  [M_i, p_j] = i\, \epsilon_{ijk} p_k, \quad [M_i, p_0] =0
\end{equation}
hold.
\newline

\noindent
{\bf 1. $\kappa$-Poincar\'e algebra in the bicrossproduct basis}.
\newline

The algebra sector reads
\begin{equation}\label{3}
   \left[N_{i}, p_{j}\right] = i\,  \delta_{ij}
 \left( {\kappa\over 2} \left(
 1 -e^{-2{p_{0}/ \kappa}}
\right) + {1\over 2\kappa} \vec{p}\,{}^{ 2}\, \right) - i\,
{1\over \kappa} p_{i}p_{j} ,
\end{equation}
and
\begin{equation}\label{4}
  \left[N_{i},p_{0}\right] = i\, p_{i}.
\end{equation}
with the first Casimir equal
\begin{equation}\label{5}
 m^2 = \left(2\kappa \sinh \left(\frac{p_0}{2\kappa}\right)\right)^2 - \vec{p}\,{}^2\, e^{p_0/\kappa}.
\end{equation}
It should be noted in passing the the parameter $m$ above is {\em
not} the  physical mass\footnote{The physical mass is defined by
equation
$\frac{1}{m_{phys}}=\lim_{p\rightarrow0}\frac{1}{p}\frac{dp_0}{dp}$,
$p =|\vec{p}|$. }, in fact, as shown in \cite{juse} the correct
expression for physical mass has the form
$$
{m^2_{phys}}=\frac{\kappa^2}{4}\, \left(1-\left(-\frac{m}{2\kappa} +
\sqrt{\frac{m^2}{4\kappa^2} +1}\right)^4\right)^2.
$$

Let us now turn to the co-algebra sector of the
$\kappa$-Poincar\'e algebra.  For our present purposes it would be
only necessary to know the co-product  for the momentum sector.
One has
\begin{eqnarray}
\displaystyle
&& \Delta(p_{i}) = p_{i}\otimes \bbbone +
e^{-{p_{0}/ \kappa}} \otimes p_{i}\, ,
\cr\cr
\displaystyle
&& \Delta(p_{0}) = p_{0}\otimes \bbbone +  \bbbone \otimes p_{0}\, ,
\label{6}
\end{eqnarray}
It is worth noticing that the bicrossproduct basis is singled out
by the  condition that the energy $p_0$ co-commutes.

The co-product is of crucial physical importance, because it makes
it possible to  construct the space-time sector and the phase
space of the theory by a step-by-step procedure.  Putting it
another way, any construction of the space-time sector is in a
sense equivalent to definition of some energy-momentum co-product,
and only the one described by eq.~(\ref{6}) has the virtue that
together with the commutational relations (\ref{1}--\ref{4}) it
furnishes a Hopf algebra. It should be stressed at this point that
had we not have this structure in our possession, we would not be
able to go beyond the energy-momentum sector.

The  general procedure of construction of the space-time
commutator algebra from energy-momentum  co-algebra
consists of the following steps \cite{maru}, \cite{crossalg}:

\begin{enumerate}
\item One defines the bracket (pairing) $<\star, \star>$ between momentum variables
$p,q$ and position variables $x,y$ in
a natural way as follows
\begin{equation}\label{8a}
 <p_\mu, x_\nu> =  -i \eta_{\mu\nu}, \quad \eta_{\mu\nu} = \mbox{diag}(-1,1,1,1).
\end{equation}
\item This bracket is to be consistent with the co-product structure in the following sense
$$
 <p, xy> = <p_{(1)}, x><p_{(2)}, y>,
$$
 \begin{equation}\label{8b}
 <pq,x> =<p, x_{(1)}><q_{(2)}, x_{(2)}>,
\end{equation}
where we use the natural notation for the co-product $$\Delta t = \sum
t_{(1)} \otimes t_{(2)}.$$ Note that by
definition $$<\bbbone, \bbbone> =1.$$ One sees immediately that
the fact that momenta commute translates to the fact that
positions co-commute
\begin{equation}\label{8c}
  \Delta x_\mu = \bbbone \otimes x_\mu + x_\mu \otimes \bbbone.
\end{equation}
Then the first equation in (\ref{8b}) along with (\ref{8a}) can be used to deduce the form of the space-time commutators.
\item  It remains only to derive the commutators (cross relations) between momenta and positions. These can be found
from the definition of the so-called Heisenberg double (see \cite{crossalg}) and read
\begin{equation}\label{8d}
 [p,x] =  x_{(1)}<p_{(1)}, x_{(2)}>p_{(2)}-xp
\end{equation}
where $xp$ above is standard multiplication.
\end{enumerate}

As an example let us apply this procedure in the case of the
bicrossproduct basis \cite{maru}, \cite{crossalg}, \cite{luno}. It
follows from (\ref{8b}) that
$$
<p_i, x_0 x_j> = -\frac1\kappa\, \delta_{ij}, \quad <p_i,  x_jx_0>=0,
$$
from which one gets
\begin{equation}\label{9a}
[x_0, x_i] = -\frac{i}\kappa\, x_i.
\end{equation}
Using  (\ref{8d}) we get the standard relations
\begin{equation}\label{9b}
[p_0, x_0] = i, \quad [p_i, x_j] = -i \, \delta_{ij}.
\end{equation}
However it turns out that this algebra contains one more
non-vanishing commutator, namely
\begin{equation}\label{9c}
 [p_i, x_0] = -\frac{i}\kappa\, p_i.
\end{equation}
Of course, the algebra (\ref{9a}--\ref{9c}) satisfies the Jacobi identity.
\newline

\noindent
\textbf{ 2. Magueijo--Smolin basis}
\newline

In the recent paper \cite{JoaoLee} Magueijo and Smolin proposed
another DSR theory, whose boost  generators were constructed as a
linear combination of the standard Lorentz generators and the
generator of dilatation (but in such a way that the algebra
(\ref{1}) holds.) In this basis the commutators of four-momenta
$P_\mu$ and boosts have the following form
\begin{equation}\label{10}
   \left[N_{i}, P_{j}\right] =  i\left( \delta_{ij}P_0 -
  {1\over \kappa} P_{i}P_{j} \right),
\end{equation}
and
\begin{equation}\label{11}
  \left[N_{i},P_{0}\right] = i\, \left( 1 - {P_0\over \kappa}\right)\,P_{i}.
\end{equation}
It is easy to check that the Casimir for this algebra has the form
\begin{equation}\label{12}
 M^2 = \frac{P_{0}^2 - \vec{P}{}^2}{\left(1- \frac{P_0}\kappa\right)^2},
\end{equation}
and that $M$ above is the physical mass.

One
easily checks that the relation between variables $P_\mu$ and
$p_\mu$ of bicrossproduct basis is given by
\begin{equation}\label{13}
p_{i} = P_{i}
\end{equation}
$$
p_0 = - \frac\kappa2\log\left(1 - \frac{2P_0}{\kappa} + \frac{\vec{P}{}^2}{\kappa^2}\right),
$$
\begin{equation}\label{14}
 P_0 = \frac\kappa2\,\left(1 -  e^{-2p_0/\kappa} + \frac{\vec{p}\,{}^2}{\kappa^2}\right).
\end{equation}

 Using  formulas above one can without difficulty promote this algebra to the quantum algebra. This
 amounts only in using the homomorphisms (\ref{13}), (\ref{14}) to define the new co-products. They read
\begin{equation}\label{6s}
  \triangle(P_{i})=P_{i} \otimes 1 + \left(1 - \frac{2P_{0}}{\kappa} +
   \frac{\vec{P}^{2}}{\kappa^{2}}\right)^{{1}/{2}} \otimes P_{i}
\end{equation}
$$
  \triangle(P_{0})=P_{0} \otimes 1 + 1 \otimes P_{0} -
  \frac{2}{\kappa}P_{0} \otimes P_{0} + \frac{1}{\kappa^{2}}
   \vec{P}^{2}\otimes P_{0} +
$$
\begin{equation}\label{7s}
   + \frac{1}{\kappa} \left(1 - \frac{2P_{0}}{\kappa} +
   \frac{\vec{P}^{2}}{\kappa^{2}}\right)^{{1}/{2}} \, \sum P_{i}\otimes P_{i}
\end{equation}
To find the non-commutative structure of space time in Magueijo--Smolin basis we start again with eq.~(\ref{8a})
$$
 <P_\mu, X_\nu> =  -i \eta_{\mu\nu}.
$$
Let us now turn to the next step, eq.~(\ref{8b}). It is easy to
see that the only terms in (\ref{6s}), \ref{7s}),  which are
relevant for our computations are the bilinear ones, so we can
write
$$
  \triangle(P_{i})=\bbbone \otimes P_{i} + P_{i} \otimes \bbbone {-\frac{1}{\kappa}}\, P_{0} \otimes P_{i} + \ldots
$$
$$
  \triangle(P_{0})= \bbbone \otimes P_{0} + P_{0} \otimes \bbbone -
  \frac{2}{\kappa}P_{0} \otimes P_{0} +  \frac{1}{\kappa}\sum P_{i}\otimes P_{i}+ \ldots
$$
It follows immediately that the only non-vanishing commutators in the position sector are
\begin{equation}\label{15}
  [X_0, X_i] = -\frac{i}\kappa\, X_i.
\end{equation}
Now we can use eq.~(\ref{8d}) to derive the form of the remaining
commutators. Since this computation is a bit tricky, let us
present the necessary steps.
$$
[P_0, X_i] = \sum_j \left<\frac{1}{\kappa} \sqrt{1 -
\frac{2P_{0}}{\kappa} +
   \frac{\vec{P}^{2}}{\kappa^{2}}} \,\,  P_{j}, X_i\right>\, P_j  + $$ $$ +
X_i <\bbbone, \bbbone>\, P_0 - X_i P_0 =
$$
$$
=\frac{1}{\kappa}\,\sum_j <P_{j}, X_i>\, P_j
$$
(we made use of the fact that the only terms linear in momenta
have non-vanishing bracket with positions) from which it follows
immediately that
\begin{equation}\label{16b}
 [P_0, X_i] =  -\frac{i}\kappa P_i
\end{equation}
and by employing the same procedure we obtain the remaining
commutators
\begin{equation}\label{16a}
 [P_0, X_0] = i\left(1 - \frac{2P_0}\kappa \right)
\end{equation}
\begin{equation}\label{16c}
  [P_i, X_j] = -i \, \delta_{ij}
\end{equation}
\begin{equation}\label{16d}
 [P_i, X_0] = -\frac{i}\kappa\, P_i.
\end{equation}
Of course, as it is easy to check, the algebra above satisfies the
Jacobi identity.

Let us note in passing an interesting property of the commutator
(\ref{16a}) namely  that for states with energy $\kappa/2$ $P_0$
and $X_0$ commute. However since it is our view that one should
not treat any particular energy-momentum algebra as a physical one
unless there are appealing physical arguments to think otherwise,
we will not dwell on this observation.
\newline

\noindent
\textbf{ 3. The classical basis}
\newline

There is yet another basis which we will present here for
comparison (this basis was first described in \cite{maslanka}; see also
\cite{kolumaso}, \cite{lukclas}, \cite{luruza}.) In this basis,
which we call the classical one, the boosts--momenta commutators
together with the Lorentz sector form the classical Poincar\'e
algebra, to wit
\begin{equation}\label{17}
   \left[N_{i}, {\cal P}_{j}\right] = i\, \delta_{ij}\, {\cal P}_0  ,\quad \left[N_{i}, {\cal P}_{0}\right] = i\,
   {\cal P}_i.
\end{equation}
The Casimir for this basis equals, of course the one of special relativity, to wit
\begin{equation}\label{17a}
 {\cal M}^2 = {\cal P}_0^2 - \vec{\cal P}\,{}^2
\end{equation}
The classical generators ${\cal P}_\mu$ are related to the bicrossproduct basis generators by the formulas
\begin{equation}\label{18}
 {\cal P}_{0} = \kappa \sinh\frac{p_0}\kappa + e^{p_0/\kappa}\, \frac{\vec{p}\, {}^2}{2\kappa},
\end{equation}
\begin{equation}\label{19}
 {\cal P}_{i}= e^{p_0/\kappa} \, p_i
\end{equation}
and one can easily compute the expression for co-product
$$
  \Delta({\cal P}_{0}) = \frac\kappa2\left(K \otimes K - K^{-1}\otimes K^{-1}\right)+$$
  \begin{equation}\label{20}
  + \frac1{2\kappa}\left(K^{-1} \vec{{\cal P}}{}^2 \otimes K +
  2K^{-1} {\cal P}_i \otimes {\cal P}_i + K^{-1}\otimes K^{-1}\vec{{\cal P}}{}^2\right),
\end{equation}
\begin{equation}\label{20a}
 \Delta({\cal P}_{i}) ={\cal P}_{i}\otimes K + \bbbone\otimes {\cal P}_{i}
\end{equation}
where
$$
K = e^{p_0/\kappa} = \frac1\kappa\, \left[ {\cal P}_{0} + \left(
{\cal P}_{0}^2 - \vec{{\cal P}}{}^2  +
\kappa^2\right)^{1/2}\right].
$$

To derive the phase space commutators, we again start with the duality relation
$$
 <{\cal P}_\mu, {\cal X}_\nu> =  -i \eta_{\mu\nu},
$$
and to get the commutators in the position sector as above we take
 the part of the co-product up to the bilinear terms
\begin{equation}\label{cpcb1}
  \triangle({\cal P}_{i})=\bbbone \otimes {\cal P}_{i} + {\cal P}_{i} \otimes \bbbone +
   {\frac{1}{\kappa}}\, {\cal P}_{i} \otimes {\cal P}_{0} + \ldots
\end{equation}
\begin{equation}\label{cpcb2}
  \triangle({\cal P}_{0})= \bbbone \otimes {\cal P}_{0} + {\cal P}_{0} \otimes \bbbone +
  \frac{1}{\kappa}\sum {\cal P}_{i}\otimes {\cal P}_{i}+ \ldots
\end{equation}
which leads again to
\begin{equation}\label{21}
  [{\cal X}_0, {\cal X}_i] = -\frac{i}\kappa\, {\cal X}_i.
\end{equation}
Then by employing the same method as in the preceding subsection
we find the space-time commutators
\begin{equation}\label{22a}
 [{\cal P}_0, {\cal X}_0] = \frac{i}2\left(K + K^{-1}- \frac1{\kappa^2}\vec{{\cal
    P}}{}^2\, K^{-1}\right)
\end{equation}
\begin{equation}\label{22b}
 [{\cal P}_0, {\cal X}_i] =  -\frac{i}\kappa {\cal P}_i
\end{equation}
\begin{equation}\label{22c}
  [{\cal P}_i, {\cal X}_j] = - i \, K\delta_{ij}
\end{equation}
\begin{equation}\label{22d}
 [{\cal P}_i, {\cal X}_0] = 0.
\end{equation}
From these three examples of $\kappa$-DSR algebras, we see that
the number of different  algebras of this type is unlimited. Even
worse, if we are to treat them seriously, we see that they differ
by fundamental physical predictions like the existence of maximal
momentum and/or energy, the form of dispersion relation etc. They
share however, quite unexpectedly, one feature, namely the form of
space-time non-commutativity. Keeping this in mind, let us now
turn to considerations of more general character.

\section{General $\kappa$-DSR algebra}

Consider a generic algebra satisfying the conditions (a)--(d)
above. The Lorentz sector of such algebra is  given by
eq.~(\ref{1}). The most general expression for commutators of
boosts and momenta, compatible with rotational symmetry takes the
form
\begin{equation}\label{23a}
 [N_i, p_j] = \delta_{ij} A + p_i p_j B + \epsilon_{ijk} p_k D,
\end{equation}
\begin{equation}\label{23b}
  [N_i, p_0] = C p_i,
\end{equation}
where $A,B,C,D$ are functions of $p_0$ and $\vec{p}\,{}^2$ only
and in the limit $\kappa\rightarrow\infty$, $A$  and $C$ go to $1$
and $B$ and $D$ go to $0$. One easily sees that the Jacobi
identity for the commutator (\ref{23b})  forces the function $D
=0$, and the one for (\ref{23a}) gives the following condition for
the remaining three functions
\begin{equation}\label{24}
 \frac{\partial A}{\partial p_0} \, C + 2 \frac{\partial A}{\partial \vec{p}\,{}^2}
 \left(A + \vec{p}\,{}^2 \, B\right) - AB = 1.
\end{equation}
This condition can be solved for example for $B$ and we see that
only two of the functions  $A$, $B$, $C$ are independent.
Recalling now that the most general transformation from the
classical basis to another one satisfying the condition of
rotational covariance has the form
\begin{equation}\label{25}
p_i = \alpha({\cal P}_0, \vec{{\cal    P}}{}^2){\cal P}_i, \quad p_0 = \beta({\cal P}_0, \vec{{\cal    P}}{}^2),
\end{equation}
and is defined by two arbitrary functions $\alpha$ and $\beta$,
and that, after substituting  to eqs.~(\ref{17}) one obtains
equations of the form (\ref{23a}), (\ref{23b}), we see that any
DSR basis can be obtained from the classical basis in this way.

One should remember however that the functions $\alpha$ and
$\beta$ are not completely arbitrary.  The important physical
requirement is that any basis has the standard Poincar\'e algebra
as its limit at $\kappa\rightarrow\infty$. This means that, for
large $\kappa$ we have
$$
 \alpha({\cal P}_0, \vec{{\cal    P}}{}^2) \sim 1 + a \frac{{\cal P}_0}{\kappa} + O\left(\frac{1}{\kappa^2}\right),
$$
\begin{equation}\label{26}
\beta({\cal P}_0, \vec{{\cal    P}}{}^2) \sim {\cal P}_0\left(1 + b \frac{{\cal P}_0}{\kappa} +
O\left(\frac{1}{\kappa^2}\right)\right),
\end{equation}
where $a$, $b$ are numerical parameters, from which it follows that
$$
 {\cal P}_i \sim \left(1 - a \frac{ p_0}{\kappa} + O\left(\frac{1}{\kappa^2}\right)\right)p_i,
$$
\begin{equation}\label{27}
{\cal P}_0 \sim \left(1 - b \frac{p_0}{\kappa} + O\left(\frac{1}{\kappa^2}\right)\right)p_0.
\end{equation}
 This observation has important consequences. To see why, let us consider the commutator $[x_0, x_i]$
 in the basis defined by eqs.~(\ref{25}).  The first step is to find the form of the co-product for
 $p_i$, $p_0$ in the new basis. Using eqs.~(\ref{cpcb1}), (\ref{cpcb2}), and (\ref{27}) we have
$$
\Delta(p_i) = \bbbone \otimes p_{i} + p_{i} \otimes \bbbone +
  {\frac{a}{\kappa}}\, p_{0} \otimes p_{i} + {\frac{1+a}{\kappa}}\, p_{i} \otimes p_{0} + \ldots
$$
$$
\Delta(p_0) =\bbbone \otimes p_{0} + p_{0} \otimes \bbbone +
  \frac{1}{\kappa}\sum p_{i}\otimes p_{i}+ \ldots
$$
where $\ldots$ denote  terms that do not contribute to the
commutator. Following our prescription of the preceding  section
we find that
$$
<p_i, x_0 x_j> = \frac{a}\kappa\, \delta_{ij}, \quad <p_i,  x_jx_0>=\frac{1+a}\kappa\, \delta_{ij},
$$
from which one gets
\begin{equation}\label{28}
[x_0, x_i] = -\frac{i}\kappa\, x_i.
\end{equation}
This shows that the form of space--time non-commutativity is a basis-independent property of any $\kappa$-DSR algebra.

\section{Space-time Lorentz transformations}

Having obtained the general form of $\kappa$-DSR algebra, let us
turn to the next problem. In this section  we derive the
transformation laws that govern the action of boosts on space-time
variables. There are two equivalent ways to find this. One can
make use of the Jacobi identity for the algebra consisting of
Lorentz generators, commutators of boosts with momenta and the
phase space algebra to derive the commutators $[N_i, x_\mu]$. In
other words one assumes that the phase space commutators transform
covariantly under boosts. There is another way to get the same
result, however. To do that one takes the co-product for boosts
and proceeds following exactly the same steps as the ones that led
us to phase space algebra. In this way the Jacobi identity for the
whole of the algebra (including the commutators of boosts with
space time variables) is  guaranteed by construction.
\newline

Let us start with co-product for the boosts in the bicrossproduct basis
\begin{equation}\label{29}
 \Delta(N_{i}) = N_{i}\otimes \bbbone  +
e^{-{p_{0}/ \kappa}}\otimes N_{i} + {1\over \kappa}
\epsilon_{ijk}p_{j}\otimes M_{k}
\end{equation}
along with the natural definition for pairing
\begin{equation}\label{30}
<N_i, x_j> = i \delta_{ij} x_0, \quad <N_i, x_0> = i  x_i.
\end{equation}
Then, after simple computations, which follow the steps leading to
eq.~(\ref{9b}), (\ref{9c})  we get \cite{luno}, \cite{lukps}
\begin{equation}\label{31}
[N_i, x_j] = i \delta_{ij} x_0 - \frac{i}\kappa\, \epsilon_{ijk} M_k, \quad [N_i, x_0] = i x_{i} - \frac{i}\kappa\, N_i.
\end{equation}
One can easily check that this algebra together with (\ref{10}),
(\ref{11}), (\ref{9a} - \ref{9c})  satisfies Jacobi identity. One
can write the commutators (\ref{31}) in a slightly more convenient
way by using the following representation for rotation and boost
generators
$$
M_i = \epsilon_{ijk}x_jp_k, \;\; N_i =x_i \left(\frac\kappa2 \left(1-e^{-2p_0/\kappa}\right) +
\frac{\vec{p}\,{}^2}{2\kappa}\right) - x_0 p_i,
$$
to get
$$
[N_i, x_j] = i \delta_{ij} x_0 - \frac{i}\kappa\, x_i p_j + \frac{i}\kappa\, x_j p_i,
$$
\begin{equation}\label{32}
 [N_i, x_0] =  i x_i \left(\frac12 \left(1+e^{-2p_0/\kappa}\right) -\frac{\vec{p}\,{}^2}{2\kappa^2}\right)
 + \frac{i}\kappa\,x_0 p_i.
\end{equation}
It is easy to see that the expression (\ref{31}) is basis
independent. To check this, consider any  basis $(P_0, P_i, X_0,
X_i)$ related to the bicrossproduct one by
$$
p_i = \alpha(P_0, \vec{P}\,{}^2) P_i, \quad p_0 = \beta(P_0, \vec{P}\,{}^2),
$$
where the functions $\alpha$ and $\beta$ again have the expansion
$$
 \alpha({ P}_0, \vec{{    P}}{}^2) \sim 1 + a \frac{{ P}_0}{\kappa} + O\left(\frac{1}{\kappa^2}\right),
$$
\begin{equation}\label{33}
\beta({ P}_0, \vec{{    P}}{}^2) \sim { P}_0\left(1 + b \frac{{ P}_0}{\kappa} +
O\left(\frac{1}{\kappa^2}\right)\right).
\end{equation}
Taking now the pairing as above,
$$
<N_i, X_j> = i \delta_{ij} X_0, \quad <N_i, X_0> = i  X_i, $$ $$ <P_\mu, X_\nu> = -i \eta_{\mu\nu},
$$
we get
$$
[N_i, X_j] =$$ $$ \bbbone\, < N_i, X_j >\, \bbbone +
\frac1\kappa\, \epsilon_{ikl} <\alpha({ P}_0, \vec{{    P}}{}^2)\, P_k, X_j >\, M_l =
$$
\begin{equation}\label{34}
 = i \delta_{ij} X_0 - \frac{i}\kappa\, \epsilon_{ijk} M_k.
\end{equation}
Similarly
$$
[N_i, X_0] = \bbbone\, < N_i, X_0 >\, \bbbone +
\bbbone\, <e^{-\beta({ P}_0, \vec{{    P}}{}^2)/\kappa}, X_0 >\, N_i =
$$
\begin{equation}\label{35}
 = i  X_i - \frac{i}\kappa\,  N_i.
\end{equation}
Let us note that by using an appropriate form of $M_i$, $N_i$
generators expressed in terms of $X_\mu$,  $P_\mu$ one can obtain
the result similar to (\ref{32}) in any basis.

The next natural question to be asked is what is a natural form,
which replaces the Minkowski metric.  To find it let us observe
that by replacing $X_i$ with
\begin{equation}\label{tx}
\tilde{X}_i = X_i - \frac1\kappa N_i,
\end{equation}
 the algebra (\ref{34}), (\ref{35}) takes the simple form
\begin{equation}\label{39}
[N_i, \tilde{X}_j] = i X_0, \quad [N_i, X_0]= i \tilde{X}_j,
\end{equation}
from which it follows that the invariant quadratic form is
\begin{equation}\label{40}
X_0^2 - \tilde{X}_i\tilde{X}_i = X_0^2 -\left(X_i - \frac1\kappa N_i\right)\left(X_i - \frac1\kappa N_i\right).
\end{equation}
This quadratic form is again universal, and invariant for any
basis.

\section{Intermezzo: So where are we?}

Let us pause for a moment to summarize what we have achieved so
far. We started from a list of  energy-momentum quantum algebras
and by making use of their co-product structure we derived the
basis-independent space-time commutators.  By using the same
method we found that the form of commutator of boosts with
positions is also basis-independent. This means that the whole
space-time sector of $\kappa$-DSR is unambiguous. Instead of using
the $X_i$ position variables following from the Heisenberg double
procedure \cite{luno}, \cite{lukps}, we find it more convenient to
use the combination $\tilde{X}_i$ defined by (\ref{tx}). In terms
of these variables space-time of DSR is described by the following
commutators (because we are now in the space-time sector, it is
reasonable to replace the deformation parameter $\kappa$ with the
parameter $\ell$ of dimension of length.)
$$
[M_i, M_j] = i\, \epsilon_{ijk} M_k, \quad [M_i, N_j] = i\, \epsilon_{ijk} N_k,
$$
\begin{equation}\label{41}
  [N_i, N_j] = -i\, \epsilon_{ijk} M_k.
\end{equation}
\begin{equation}\label{42}
 [X_0, \tilde{X}_i] = - i\ell^2 \, N_i, \quad [\tilde{X}_i, \tilde{X}_j] =  i\ell^2 \,\epsilon_{ijk} M_k.
\end{equation}
\begin{equation}\label{43}
[N_i, \tilde{X}_j] = i X_0, \quad [N_i, X_0]= i \tilde{X}_j,
\end{equation}
\begin{equation}\label{43a}
[M_i, \tilde{X}_j] = i\, \epsilon_{ijk} \tilde{X}_k , \quad [M_i, X_0]= 0,
\end{equation}
It is a remarkable fact that this algebra provides the description
of non-commutative Minkowski space  (cf., the invariant element
(\ref{40})) enjoying ordinary Lorentz symmetry, and is identical
with the algebra postulated by Snyder in his seminal paper
\cite{snyder}, where the idea of space-time non-commutativity was
contemplated for the first time.  From our perspective this
algebra is exactly the universal basis of the Doubly Special
Relativity theory, we have been looking for, and thus this is the
algebra (\ref{41}--\ref{43a}) which is to be taken as a starting
point in construction of the DSR theory and analysis of its
physical predictions. Specifically, this algebra furnishes the
structure of configuration space of the DSR and thus can be used
to construct the second order particle lagrangian. This in turn
will make it possible to define physical four-momenta defined by
particle dynamics and not the formal algebraic manipulations.
However, it is still of interest to investigate the natural phase
space basis associated with the space-time algebra above.

\section{The Snyder's basis}

Promoting (\ref{41}--\ref{43}) to the  universal space-time
algebra brings the problem of construction of  Doubly Special
Relativity into completely new perspective. In short we would like
to find a representation of this algebra on the space of functions
of four commuting variables, which can be identified with
components of momenta. There are, of course many such
representations, which correspond to different energy momentum
bases discussed above. However there is clearly the most natural
basis, namely the one considered by Snyder. In this basis the
Lorentz generators are given by classical formulas, to wit
\begin{equation}\label{44}
 M_i = \epsilon_{ijk}  \tilde{X}_j P_k , \quad N_i = \left( \tilde{X}_i P_0 - {X}_0 P_i\right)
\end{equation}
from which one can derive the form of the commutators $[P_\mu, X_\nu]$ as follows.
\begin{equation}\label{45}
 [P_i, \tilde{X}_j] = - i \delta_{ij} - i \ell^2 P_i P_j,
\end{equation}
\begin{equation}\label{46}
  [P_0, X_0] = i\left(1 - \ell^2 P_0^2\right),
\end{equation}
\begin{equation}\label{47}
 [P_i, X_0] = -[\tilde{X}_i, P_0] = - i \ell^2 P_i P_0.
\end{equation}
It can be easily checked that the variables $P_\mu$ transform under rotations and boosts in a classical way, to wit
\begin{equation}\label{48}
[M_i, P_j]= i\epsilon_{ijk}P_k, \quad [M_i, P_0]=0,
\end{equation}
\begin{equation}\label{49}
  [N_i, P_j]= i \delta_{ij}P_0, \quad [N_i, P_0]=i P_i.
\end{equation}
Thus the Snyder's basis is a different, and certainly more
convenient realization of a basis with classical  energy-momentum
algebra. Indeed, returning to the generators $X_i = \tilde{X}_i +
\ell N_i$ we find the relations
\begin{equation}\label{50}
 [P_i, {X}_j] = - i \delta_{ij}\left(1 + \ell P_0\right) - i \ell^2 P_i P_j,
\end{equation}
\begin{equation}\label{51}
  [{X}_i, P_0] =  i\ell P_i + i \ell^2 P_i P_0,
\end{equation}
as compared with the non-polynomial relations (\ref{22a} --
\ref{22d}). Of course, one can work out the  co-product structure
for the algebra (\ref{48}), (\ref{49}) resulting in relations
(\ref{50}), (\ref{51})

It is a remarkable fact that the algebra (\ref{45}) once reduced
to one space dimension is exactly the  one considered by many as
an example of deformed commutational relations leading to
generalized uncertainty relations indicating existence of the
minimal length. The detailed discussion of quantum mechanical
structure underlying this uncertainty relation can be found in
\cite{kmm} and we will not repeat it here. One should note however
that the extension to three dimension  presented in this paper
differs from ours.

We see therefore that the algebra (\ref{44} -- \ref{47}) predicts
the existence of minimal  length equal $\ell$.  The question
arises as to if this minimal length is  observer-independent, as
required by DSR (the speed o light is of course
observer-independent because the dispersion relation is
classical.) By this we mean the following. Suppose we have two
inertial observers attempting do find out what is their value of
the minimal length associated with their generalized uncertainty
relation. For both of them this minimal length is associated with
the value of the parameter $\ell$ in their commutational
relations. The statement that the minimal length is
observer-independent is equivalent therefore with the statement
that the parameter $\ell$ is the same in any inertial frame. To
see that this is indeed the case let us consider the infinitesimal
boosts parameterized by parameter $\varepsilon^k$
$$
\tilde{X}_i' = \tilde{X}_i + \frac1i\,\varepsilon^k\, [N_k,
\tilde{X}_i], \quad P_i' = P_i +  \frac1i\,\varepsilon^k\, [N_k,
P_i]
$$
and consider the commutator $[P'_i, \tilde{X}_j']$ to order $\varepsilon$.
One can easily check that to this order
\begin{equation}\label{52}
 [P'_i, \tilde{X}_j'] = - i \delta_{ij} - i \ell^2 P'_i P'_j,
\end{equation}
which is the desired result. It should be stressed that to get
this result one should make use of  eq.~(\ref{47}) in a nontrivial
way.

\section{Conclusions and outlook}

Let us conclude this paper with an outline of the logic of
derivations presented above. We started  with a simple observation
that if one wants to base the DSR theory on the energy-momentum
sector, one encounters immediately the problem as to which of
mathematically equivalent bases of  $\kappa$-Poincar\'e algebra
should be taken as a starting point. There are clearly two
possibilities: either one of the bases is distinguished or one
should look for some basis independent elements of the DSR theory.
The next observation is that  the sole energy-momentum sector is
insufficient to construct a physical theory; one needs the whole
phase space.

At this point the quantum structure of $\kappa$-Poincar\'e algebra
turns out to be crucial.  It is only for existence of co-product
that one can construct the space-time sector is an unambiguous
way. It should be stressed at this point that this provides the
physical interpretation of co-product in the case of algebra of
space-time symmetries. Thus the co-product of momenta is the tool
enabling the construction of space-time.

Having constructed the phase space, we observe that the structure
of space-time sector, i.e.,  the space-time non-commutativity as
well as the action of boosts and rotations on space-time variables
is actually completely independent of the energy-momentum basis we
started with. We have therefore a non-commutative space-time,
which might be, like space-time of Special Relativity, an arena
for physical phenomena. It seems  reasonable to take this
space-time structure as a starting point of investigation of the
DSR theory. One should stress at this point that the uniqueness is
not the only virtue of the space-time algebra derived in this
paper. First of all, when the space-time non-commutativity is
expressed in terms of variables $\tilde{X}_i$ on obtains the
algebra (\ref{42}) being the simplest possible extension of the
non-commutative algebra motivated by string theory (for recent
review see \cite{douglasnekrasov})
\begin{equation}\label{53}
  [X_\mu, X_\nu] = i \Theta_{\mu\nu},
\end{equation}
where $\Theta_{\mu\nu}$ is central element, which  (contrary to
(\ref{53})) manifestly preserves  Lorentz invariance. Second if
following Snyder \cite{snyder} one takes a natural representation
of the space-time algebra one discovers the natural appearance of
observer-independent length scale $\ell$. At this point we would
decline from any strong statement as to if this particular phase
space basis is physical , i.e., if the variables $P_\mu$ of
Snyder's basis are to be identified with physical momenta, though
certainly this basis looks very promising in this respect. Third
the construction presented above provides a non-commutative
Minkowski space with undeformed Lorentz symmetry.

There is of course a lot of open problems and possible areas of
future research.  It seems to us that the most urgent one is to
try to construct a natural particle dynamics on non-commutative
space-time. This would enable one to identify physical momentum
associated with particle motion. Another approach to the same
problem would be to formulate a field theory on our
non-commutative space-time along the lines proposed long ago in
\cite{snyder2}. Quantization of such theory would also most likely
make it possible to find out what the physical momentum is. Last
but not least the fact that along with a unique space-time we seem
to have plethora of energy-momentum spaces certainly requires
explanation.

\section*{Acknowledgement}

We would like to thank J.~Lukierski for discussions and comments on early version of the manuscript.



\begin{thebibliography}{99}
\bibitem{gacphen} G.~Amelino-Camelia, {\tt gr-qc/0204051} and references therein.
\bibitem{gac1} G.~Amelino-Camelia,  Int.~J.~Mod. Phys. {\bf D 11}, 35 (2002), [{\tt gr-qc/0012051}].
\bibitem{gac2} G.~Amelino-Camelia,  Phys.~Lett. {\bf B 510}, 255, (2001), [{\tt hep-th/0012238}].
\bibitem{gacnew} G.~Amelino-Camelia, {\tt gr-qc/0106004}, Proceedings of the 37th
Karpacz Winter School on Theoretical Physics, to appear.
\bibitem{jkgminl} J. Kowalski-Glikman, Phys.~Lett. {\bf A 286}, 391 (2001), [{\tt hep-th/0102098}].
\bibitem{rbgacjkg} N.R.~Bruno, G.~Amelino-Camelia, and J. Kowalski-Glikman, Phys.~Lett. {\bf B 522},
133 (2001), [{\tt hep-th/0107039}].
\bibitem{lunoruto} J.~Lukierski, A.~Nowicki, H.~Ruegg and V.N.~Tolstoy,
Phys. Lett.  {\bf B264}, 331 (1991).
\bibitem{maru} S. Majid and H. Ruegg, Phys. Lett. {\bf B334},
348 (1994).
\bibitem{snyder} H.S.~Snyder, Phys.~Rev. {\bf 71}, 38 (1947).
\bibitem{rurev} H.~Ruegg in {\em Integrable Systems, Quantum Groups, and Quantum Field Theories},
L.A.~Ibort and M.A.~Rodr\'iguez eds., NATO ASI series, vol.~409,
Kluwer Academic Publishers, Dordrecht, 1993; J.~Lukierski,
 H.~Ruegg and V.N.~Tolstoy, in {\em Quantum Groups: Formalism and
 Applications}, Polish Scientific Publishers, (1995), p. 259.
\bibitem{kolumaso} P.~Kosi\'nski, J.~Lukierski, P.~Maslanka, J.~Sobczyk, Mod.~Phys.~Lett. {\bf A10}, 2599 (1995).
\bibitem{JoaoLee} J.~Magueijo and L.~Smolin, {\tt hep-th/0112090}.
\bibitem{juse} J.~Kowalski-Glikman and S.~Nowak, {\tt hep-th/0203040}.
\bibitem{lunoDSR} J. Lukierski and A. Nowicki, {\tt hep-th/0203065}.
\bibitem{luno} J. Lukierski and A. Nowicki, Proceedings of
Quantum Group Symposium at Group 21, (July 1996, Goslar) Eds.
H.-D. Doebner and V.K. Dobrev, Heron Press, Sofia, 1997, p.
186.
\bibitem{crossalg}  A.~Nowicki, in Proceedings of IX Max Born Symposium {\em New Symmetries in the
Theories of Fundamental Interactions}, Polish Scientific
Publishers, (1997), p. 43 [{\tt math.QA/9803064}].
\bibitem{maslanka} P.~Ma\'slanka, J.~Math.~Phys. {\bf 34}, 6025 (1993)
\bibitem{lukclas} J.~Lukierski, Proceedings of
Quantum Group Symposium at Group 21, (July 1996, Goslar) Eds.
H.-D. Doebner and V.K. Dobrev, Heron Press, Sofia, 1997, p.~173.
\bibitem{luruza} J. Lukierski, H. Ruegg and W.J. Zakrzewski, Ann.
Phys. {\bf 243}, 90 (1995).
\bibitem{lukps} J.~Lukierski, Proceedings of
the III International Workshop on Classical and Quantum Integrable
Systems, JINR (1998), [{\tt hep-th/9812063}].
\bibitem{kmm} A.~Kempf, G.~Mangano, R.~Mann, Phys.~Rev. {\bf D52}, 1108 (1995) [{\tt hep-th/9412167}];
A.~Kempf, G.~Mangano, Phys.~Rev. {\bf D55}, 1108 (1997) [{\tt hep-th/9612084}].
\bibitem{douglasnekrasov} M.R.~Douglas and N.A.~Nekrasov, Rev.~Mod.~Phys. {\bf 73}, 977 (2002) [{\tt hep/th-0106048}]
\bibitem{snyder2} H.S.~Snyder, Phys.~Rev. {\bf 71}, 68 (1947).
\end{thebibliography}
\end{document}